\documentclass[aps,prl,reprint,superscriptaddress,amsmath,amssymb,floatfix,showkeys,showpacs]{revtex4-2}
\pdfoutput=1
\usepackage{bm}
\usepackage{esint} 
\usepackage[normalem]{ulem}
\usepackage{graphicx}
\usepackage{natbib}
\usepackage[colorlinks=true,linkcolor=blue,citecolor=blue,urlcolor=blue]{hyperref}
\usepackage{pr_macros}
\usepackage{xcolor}

\newcommand{\dg}{\delta_{g}}
\newcommand{\dm}{\delta_{m}}
\newcommand{\tdg}{\tilde{\delta}_{g}}
\newcommand{\qpar}{Q_{\mathrm{sat}}}
\newcommand{\qparres}{\qpar{=}0.05{\pm}0.14}
\newcommand{\mham}{M_h^{\mathrm{AM}}}

\newcommand{\ngal}{N_{\mathrm{gal}}}
\newcommand{\ngalmh}{N_{\mathrm{gal}}(M_h)}
\newcommand{\ngalmhx}{N_{\mathrm{gal}}(M_h | \mathbf{X})}
\newcommand{\ngalmhdg}{N_{\mathrm{gal}}(M_h | \dg)}
\newcommand{\nsat}{N_{\mathrm{sat}}}
\newcommand{\nsatmhdg}{N_{\mathrm{sat}}(M_h | \dg)}
\newcommand{\hmpc}{h^{-1}\,\mathrm{Mpc}}
\newcommand{\mpch}{\mathrm{Mpc}/h}
\newcommand{\hmsol}{h^{-1}M_{\odot}}
\newcommand{\dsh}{\Delta\Sigma}

\newcommand{\wphg}{w_p^{hg}}
\newcommand{\xihg}{\xi_{hg}}
\newcommand{\xihm}{\xi_{hm}}
\newcommand{\xihx}{\xi_{hx}}
\newcommand{\rs}{r_{s}}

\begin{document}

\preprint{APS/123-QED}
\title{Direct Measurement of Galaxy Assembly Bias using DESI DR1 Data}

\author{Z.~Shao}
\affiliation{Department of Astronomy, School of Physics and Astronomy, Shanghai Jiao Tong University, Shanghai 200240, China}
\affiliation{State Key Laboratory of Dark Matter Physics, \& Tsung-Dao Lee Institute, Shanghai Jiao Tong University, Shanghai 200240, China}

\author{Y.~Zu}
\email[]{yingzu@sjtu.edu.cn}
\affiliation{Department of Astronomy, School of Physics and Astronomy, Shanghai Jiao Tong University, Shanghai 200240, China}
\affiliation{State Key Laboratory of Dark Matter Physics, \& Tsung-Dao Lee Institute, Shanghai Jiao Tong University, Shanghai 200240, China}

\author{A.~N.~Salcedo}
\affiliation{Department of Astronomy/Steward Observatory, University of Arizona, 933 North Cherry Avenue, Tucson, AZ 85721-0065, USA}
\affiliation{Department of Physics, University of Arizona, 1118 East Fourth Street, Tucson, AZ 85721, USA}

\author{J.~Wang}
\affiliation{Department of Astronomy, School of Physics and Astronomy, Shanghai Jiao Tong University, Shanghai 200240, China}

\author{X.~Yang}
\affiliation{Department of Astronomy, School of Physics and Astronomy, Shanghai Jiao Tong University, Shanghai 200240, China}
\affiliation{State Key Laboratory of Dark Matter Physics, \& Tsung-Dao Lee Institute, Shanghai Jiao Tong University, Shanghai 200240, China}

\author{D.~H.~Weinberg}
\affiliation{Department of Astronomy and Center for Cosmology and AstroParticle Physics, The Ohio State University, Columbus, OH 43210, USA}

\author{X.~Xu}
\affiliation{Department of Astronomy, School of Physics and Astronomy, Shanghai Jiao Tong University, Shanghai 200240, China}
\affiliation{Shanghai Key Lab for Astrophysics, Shanghai Normal University, Shanghai 200234, China}

\author{Z.~Zhai}
\affiliation{Department of Astronomy, School of Physics and Astronomy, Shanghai Jiao Tong University, Shanghai 200240, China}

\author{Z.~Zhang}
\affiliation{Department of Astronomy, School of Physics and Astronomy, Shanghai Jiao Tong University, Shanghai 200240, China}
\affiliation{State Key Laboratory of Dark Matter Physics, \& Tsung-Dao Lee Institute, Shanghai Jiao Tong University, Shanghai 200240, China}

\author{J.~Aguilar}
\affiliation{Lawrence Berkeley National Laboratory, 1 Cyclotron Road, Berkeley, CA 94720, USA}

\author{S.~Ahlen}
\affiliation{Department of Physics, Boston University, 590 Commonwealth Avenue, Boston, MA 02215 USA}

\author{D.~Bianchi}
\affiliation{Dipartimento di Fisica ``Aldo Pontremoli'', Universit\`a degli Studi di Milano, Via Celoria 16, I-20133 Milano, Italy}
\affiliation{INAF-Osservatorio Astronomico di Brera, Via Brera 28, 20122 Milano, Italy}

\author{D.~Brooks}
\affiliation{Department of Physics \& Astronomy, University College London, Gower Street, London, WC1E 6BT, UK}

\author{R.~Canning}
\affiliation{Institute of Cosmology and Gravitation, University of Portsmouth, Dennis Sciama Building, Portsmouth, PO1 3FX, UK}

\author{F.~J.~Castander}
\affiliation{Institut d'Estudis Espacials de Catalunya (IEEC), c/ Esteve Terradas 1, Edifici RDIT, Campus PMT-UPC, 08860 Castelldefels, Spain}
\affiliation{Institute of Space Sciences, ICE-CSIC, Campus UAB, Carrer de Can Magrans s/n, 08913 Bellaterra, Barcelona, Spain}

\author{T.~Claybaugh}
\affiliation{Lawrence Berkeley National Laboratory, 1 Cyclotron Road, Berkeley, CA 94720, USA}

\author{S.~Cole}
\affiliation{Institute for Computational Cosmology, Department of Physics, Durham University, South Road, Durham DH1 3LE, UK}

\author{A.~Cuceu}
\affiliation{Lawrence Berkeley National Laboratory, 1 Cyclotron Road, Berkeley, CA 94720, USA}

\author{A.~de la Macorra}
\affiliation{Instituto de F\'{\i}sica, Universidad Nacional Aut\'{o}noma de M\'{e}xico,  Circuito de la Investigaci\'{o}n Cient\'{\i}fica, Ciudad Universitaria, Cd. de M\'{e}xico  C.~P.~04510,  M\'{e}xico}

\author{Arjun~Dey}
\affiliation{NSF NOIRLab, 950 N. Cherry Ave., Tucson, AZ 85719, USA}

\author{P.~Doel}
\affiliation{Department of Physics \& Astronomy, University College London, Gower Street, London, WC1E 6BT, UK}

\author{S.~Ferraro}
\affiliation{Lawrence Berkeley National Laboratory, 1 Cyclotron Road, Berkeley, CA 94720, USA}
\affiliation{University of California, Berkeley, 110 Sproul Hall \#5800 Berkeley, CA 94720, USA}

\author{J.~E.~Forero-Romero}
\affiliation{Departamento de F\'isica, Universidad de los Andes, Cra. 1 No. 18A-10, Edificio Ip, CP 111711, Bogot\'a, Colombia}
\affiliation{Observatorio Astron\'omico, Universidad de los Andes, Cra. 1 No. 18A-10, Edificio H, CP 111711 Bogot\'a, Colombia}

\author{E.~Gaztañaga}
\affiliation{Institut d'Estudis Espacials de Catalunya (IEEC), c/ Esteve Terradas 1, Edifici RDIT, Campus PMT-UPC, 08860 Castelldefels, Spain}
\affiliation{Institute of Cosmology and Gravitation, University of Portsmouth, Dennis Sciama Building, Portsmouth, PO1 3FX, UK}
\affiliation{Institute of Space Sciences, ICE-CSIC, Campus UAB, Carrer de Can Magrans s/n, 08913 Bellaterra, Barcelona, Spain}

\author{S.~Gontcho A Gontcho}
\affiliation{Lawrence Berkeley National Laboratory, 1 Cyclotron Road, Berkeley, CA 94720, USA}
\affiliation{University of Virginia, Department of Astronomy, Charlottesville, VA 22904, USA}

\author{G.~Gutierrez}
\affiliation{Fermi National Accelerator Laboratory, PO Box 500, Batavia, IL 60510, USA}

\author{K.~Honscheid}
\affiliation{Center for Cosmology and AstroParticle Physics, The Ohio State University, 191 West Woodruff Avenue, Columbus, OH 43210, USA}
\affiliation{Department of Physics, The Ohio State University, 191 West Woodruff Avenue, Columbus, OH 43210, USA}
\affiliation{The Ohio State University, Columbus, 43210 OH, USA}

\author{C.~Howlett}
\affiliation{School of Mathematics and Physics, University of Queensland, Brisbane, QLD 4072, Australia}

\author{D.~Huterer}
\affiliation{Department of Physics, University of Michigan, 450 Church Street, Ann Arbor, MI 48109, USA}
\affiliation{University of Michigan, 500 S. State Street, Ann Arbor, MI 48109, USA}

\author{M.~Ishak}
\affiliation{Department of Physics, The University of Texas at Dallas, 800 W. Campbell Rd., Richardson, TX 75080, USA}

\author{R.~Joyce}
\affiliation{NSF NOIRLab, 950 N. Cherry Ave., Tucson, AZ 85719, USA}

\author{T.~Kisner}
\affiliation{Lawrence Berkeley National Laboratory, 1 Cyclotron Road, Berkeley, CA 94720, USA}

\author{A.~Kremin}
\affiliation{Lawrence Berkeley National Laboratory, 1 Cyclotron Road, Berkeley, CA 94720, USA}

\author{O.~Lahav}
\affiliation{Department of Physics \& Astronomy, University College London, Gower Street, London, WC1E 6BT, UK}

\author{C.~Lamman}
\affiliation{The Ohio State University, Columbus, 43210 OH, USA}

\author{M.~Landriau}
\affiliation{Lawrence Berkeley National Laboratory, 1 Cyclotron Road, Berkeley, CA 94720, USA}

\author{L.~Le~Guillou}
\affiliation{Sorbonne Universit\'{e}, CNRS/IN2P3, Laboratoire de Physique Nucl\'{e}aire et de Hautes Energies (LPNHE), FR-75005 Paris, France}

\author{M.~E.~Levi}
\affiliation{Lawrence Berkeley National Laboratory, 1 Cyclotron Road, Berkeley, CA 94720, USA}

\author{M.~Manera}
\affiliation{Departament de F\'{i}sica, Serra H\'{u}nter, Universitat Aut\`{o}noma de Barcelona, 08193 Bellaterra (Barcelona), Spain}
\affiliation{Institut de F\'{i}sica d’Altes Energies (IFAE), The Barcelona Institute of Science and Technology, Edifici Cn, Campus UAB, 08193, Bellaterra (Barcelona), Spain}

\author{A.~Meisner}
\affiliation{NSF NOIRLab, 950 N. Cherry Ave., Tucson, AZ 85719, USA}

\author{R.~Miquel}
\affiliation{Instituci\'{o} Catalana de Recerca i Estudis Avan\c{c}ats, Passeig de Llu\'{\i}s Companys, 23, 08010 Barcelona, Spain}
\affiliation{Institut de F\'{i}sica d’Altes Energies (IFAE), The Barcelona Institute of Science and Technology, Edifici Cn, Campus UAB, 08193, Bellaterra (Barcelona), Spain}

\author{J.~Moustakas}
\affiliation{Department of Physics and Astronomy, Siena University, 515 Loudon Road, Loudonville, NY 12211, USA}

\author{S.~Nadathur}
\affiliation{Institute of Cosmology and Gravitation, University of Portsmouth, Dennis Sciama Building, Portsmouth, PO1 3FX, UK}

\author{N.~Palanque-Delabrouille}
\affiliation{IRFU, CEA, Universit\'{e} Paris-Saclay, F-91191 Gif-sur-Yvette, France}
\affiliation{Lawrence Berkeley National Laboratory, 1 Cyclotron Road, Berkeley, CA 94720, USA}

\author{W.~J.~Percival}
\affiliation{Department of Physics and Astronomy, University of Waterloo, 200 University Ave W, Waterloo, ON N2L 3G1, Canada}
\affiliation{Perimeter Institute for Theoretical Physics, 31 Caroline St. North, Waterloo, ON N2L 2Y5, Canada}
\affiliation{Waterloo Centre for Astrophysics, University of Waterloo, 200 University Ave W, Waterloo, ON N2L 3G1, Canada}

\author{F.~Prada}
\affiliation{Instituto de Astrof\'{i}sica de Andaluc\'{i}a (CSIC), Glorieta de la Astronom\'{i}a, s/n, E-18008 Granada, Spain}

\author{I.~P\'erez-R\`afols}
\affiliation{Departament de F\'isica, EEBE, Universitat Polit\`ecnica de Catalunya, c/Eduard Maristany 10, 08930 Barcelona, Spain}

\author{G.~Rossi}
\affiliation{Department of Physics and Astronomy, Sejong University, 209 Neungdong-ro, Gwangjin-gu, Seoul 05006, Republic of Korea}

\author{L.~Samushia}
\affiliation{Abastumani Astrophysical Observatory, Tbilisi, GE-0179, Georgia}
\affiliation{Department of Physics, Kansas State University, 116 Cardwell Hall, Manhattan, KS 66506, USA}
\affiliation{Faculty of Natural Sciences and Medicine, Ilia State University, 0194 Tbilisi, Georgia}

\author{E.~Sanchez}
\affiliation{CIEMAT, Avenida Complutense 40, E-28040 Madrid, Spain}

\author{D.~Schlegel}
\affiliation{Lawrence Berkeley National Laboratory, 1 Cyclotron Road, Berkeley, CA 94720, USA}

\author{J.~Silber}
\affiliation{Lawrence Berkeley National Laboratory, 1 Cyclotron Road, Berkeley, CA 94720, USA}

\author{D.~Sprayberry}
\affiliation{NSF NOIRLab, 950 N. Cherry Ave., Tucson, AZ 85719, USA}

\author{G.~Tarl\'{e}}
\affiliation{University of Michigan, 500 S. State Street, Ann Arbor, MI 48109, USA}

\author{B.~A.~Weaver}
\affiliation{NSF NOIRLab, 950 N. Cherry Ave., Tucson, AZ 85719, USA}

\author{R.~Zhou}
\affiliation{Lawrence Berkeley National Laboratory, 1 Cyclotron Road, Berkeley, CA 94720, USA}

\author{H.~Zou}
\affiliation{National Astronomical Observatories, Chinese Academy of Sciences, A20 Datun Road, Chaoyang District, Beijing, 100101, P.~R.~China}

\begin{abstract}
    We report the first direct measurement of galaxy assembly bias, a
    critical systematic in cosmology, from the Dark Energy Spectroscopic
    Instrument~(DESI) Bright Galaxy Survey.  We introduce a novel,
    cosmology-independent method to measure the halo occupation
    distribution~(HOD) by combining a state-of-the-art group catalog with
    weak gravitational lensing. For groups binned by total luminosity, we
    determine the galaxy occupation number $N_{\rm gal}$ from group-galaxy
    cross-correlations, while weak lensing constrains the average halo mass
    $M_h$. Applying this to a volume-limited sample at $z{\in}[0.05,0.2]$,
    we measure the dependence of HOD, $\ngal(M_h)$, on large-scale
    overdensity $\dg$.  Focusing on the satellite galaxies, we find an
    assembly bias parameter of $\qparres$, a result consistent with zero
    and in tension with many empirical galaxy formation models. Our method
    provides a robust approach for characterizing galaxy assembly bias to
    achieve precision cosmology with DESI and future Stage-V surveys.
\end{abstract}

\maketitle
\emph{Introduction}---
\label{sec:intro} Modern cosmology relies on galaxy redshift surveys to
probe the growth of dark matter structure~\citep{Weinberg2013} and the
potential time-evolution of dark energy~\citep{Adame2025DESI,DESI2025w0wa}.
The accuracy of these cosmological probes hinges on a robust model
connecting the observed galaxy population to the underlying dark matter
halos~\citep{Wechsler2018}, particularly through the number of galaxies
residing in a single halo~(i.e., the occupation number). The standard halo
occupation distribution~(HOD;
\citep{Jing1998,Peacock2000,Berlind2002,Zheng2005, Mandelbaum2006, Zu2015})
model posits that the {\it mean} occupation number $\ngal$ is a function
solely of halo mass $M_h$. However, galaxy assembly may be systematically
influenced by other halo properties. This effect, known as galaxy assembly
bias~(GAB), introduces a dependence of $\ngal(M_h)$ on secondary halo
properties $\mathbf{X}$, such as the halo formation time, concentration,
and large-scale
environment~\citep{Zhu2006,Zehavi2018,Artale2018,Bose2019,Hadzhiyska2021}.
A strong GAB is a near-ubiquitous prediction of empirical models of galaxy
formation, including semi-analytic models~\citep{Zehavi2018,Artale2018} and
hydrodynamical
simulations~\citep{Zhu2006,Bose2019,Hadzhiyska2021,Meng2025}.  As a direct
consequence, GAB has been identified as a key source of systematic
uncertainty for cosmological
inference~\citep{Findlay2025,Garcia-Quintero2025,Mena-Fernandez2025},
posing a significant risk to the fidelity of constraints from Stage-IV
surveys like the Dark Energy Spectroscopic Instrument~(DESI;
\citep{DESI2016a,DESI2016b,Guy2023DESI,Schlafly2023DESI,Miller2024DESI,Poppett2024DESI}).
In this Letter, we develop a novel method to directly measure the HOD of
group-mass halos residing in different large-scale overdensity
environments, in order to constrain the level of GAB with the DESI Data
Release 1~(DR1) data.

Over the past decade, various studies have tried to detect GAB in the local
Universe using galaxy clustering statistics~(hereafter referred to as the
``indirect method''), often assuming a fixed cosmology~(e.g.,
Refs.~\citep{Lehmann2017,Zentner2019,Vakili2019,Obuljen2020,
Yuan2021,Wang2022GAB,Salcedo2022,Pearl2024}). The key observable of galaxy
clustering is the bias factor of galaxies, $b_g$, defined as the linear
clustering amplitude of galaxies relative to the dark matter. In
particular, for a given set of cosmological parameters $\bm{\theta}$, we
can predict $b_g$ for any galaxy sample with a number density $n_g$ as:
\begin{equation}
    b_g{=}\frac{1}{n_g}\!\iint\!\! \ngalmhx
    n_h(M_h| \mathbf{X}, \bm{\theta})
    b_h(M_h | \mathbf{X},\bm{\theta})
    \, \mathrm{d}\mathbf{X}\mathrm{d} M_h,
    \label{eqn:bggab}
\end{equation}
where $\ngalmhx$ is our GAB observable, $n_h$ is the differential halo
number density distribution, and $b_h$ is the bias factor of
halos~\footnote{The term $b_h(M_h | \mathbf{X})$ describes the so-called
``halo assembly bias''
effect~\citep{Sheth2004,Gao2005,Harker2006,Wechsler2006,Jing2007,Li2008,Salcedo2018}.}.
Equation~\ref{eqn:bggab} shows that the GAB constraint using an indirect
method is susceptible to systematic uncertainties if an incorrect cosmology
is assumed for $b_h$ and $n_h$, {\it even though GAB itself does not depend on
$\bm{\theta}$}.  Conversely, cosmological constraints using galaxy
clustering may be biased if GAB is not properly modeled~\cite{Wang2025}.
This issue may be related to the lensing-clustering discrepancy of the
luminous red galaxies~(LRGs) observed at
$z{<}0.7$~\citep{Leauthaud2017,Lange2019,Yuan2020,Amon2023,Shao2023}.  To
mitigate such an issue in the Sloan Digital Sky Survey~(SDSS;
\citep{York2000}), \citet{Salcedo2022} employed the {\it simulated} halo
catalog from the \texttt{ELUCID} constrained simulation that accurately
reconstructed the initial density perturbation~(hence the correct
$n_h{\times}b_h$) within the SDSS volume~\citep{Wang2016}, and they found
no evidence for a significant GAB in the SDSS data.

However, an ideal method would directly measure $\ngalmhx$ for the {\it
observed} halos selected by secondary halo properties $\mathbf{X}$, without
the need to assume a specific cosmology.  In this work, we make the first
step towards this goal by taking advantage of the state-of-the-art DESI
halo-based group catalog developed by \citet{Yang2021}, which
achieves ${>}90\%$ completeness and purity for halos with
$M_h{>}10^{13}\,\hmsol$ at $z{<}0.5$.  More important, the halo-based group
finder is not biased towards red galaxy colors~\citep{Golden-Marx2023}.
Since GAB may influence the color distribution of satellite galaxies within
halos, a color-insensitive group finder is essential for reliable GAB
detection.  For the choice of $\mathbf{X}$, we use the large-scale galaxy
overdensity $\dg$ around each halo, which serves as a proxy for the
underlying dark matter overdensity $\dm$.  \citet{Xu2021} showed that the
GAB effect in semi-analytic models can be fully accounted for by $\dm$
enclosed within several halo radii.

To measure $\ngal$ for DESI groups binned by $\dg$, one
might initially consider simply counting member galaxies identified by the
group finder~(i.e., using ``richness''). However, accurately determining
membership probabilities remains challenging, particularly due to
interlopers caused by projection effects~\citep{Zu2017}. In this work, we
instead introduce a robust alternative approach that does not depend on the
group finder: we measure $\ngal$ using the cross-correlation between groups
and galaxies and remove the interlopers statistically. By combining these
$\ngal$ measurements with halo mass $M_h$ derived from weak lensing, we {\it directly}
measure the GAB observable $\ngalmhdg$.

Throughout this letter, we use the latest {\it Planck} $\Lambda$CDM model
for distances~\citep{Planck2020}, and adopt the halo mass definition in
which the matter density enclosed within the halo radius $r_h$ is $200$
times the mean density $\rho_m$ of the Universe.


\emph{DESI Bright Galaxy Survey and group catalog}---We use the large-scale structure~(LSS) galaxy catalog~\citep{Ross2025}
derived from the flux-limited~($m_r{<}19.5$) sample of the Bright Galaxy
Survey~(BGS; \citep{Hahn2023}) within DESI DR1~\citep{DESI2025}. The LSS
catalog provides the necessary weights to correct for imaging
systematics~($w_{\rm sys}$), fiber assignment incompleteness~($w_{\rm
comp}$), and redshift failures~($w_{\rm zfail}$), so that the final LSS
weight of each galaxy is $w_{\rm tot}{=}w_{\rm sys}w_{\rm comp}w_{\rm
zfail}$.  For our GAB detection analysis, we construct a volume-limited
galaxy sample from the LSS catalog with $M_r{\leqslant}{-}19.68$~(i.e.,
$L{\geqslant}0.4 L_*$) within $0.05<z<0.2$, yielding an average
galaxy number density of
$\bar{n}_g \approx 0.01\,h^3{\rm Mpc^{-3}}$.

For the host halos, we employ an updated version of the extended halo-based
group catalog developed in \citet{Yang2021}, which has incorporated all
spectroscopic redshifts observed by DESI DR2. To ensure a high purity above~$90\%$, we
select groups with $\log\mham{>}13.0$, where $\mham$ is the
catalog-provided halo mass estimated from abundance matching~(AM) using the
total luminosity of all the group members~($L_{\rm tot}$).  We set the
group center at the brightest cluster galaxy~(BCG) following the
practice in Ref.~\cite{Wang2022}.  However, a fraction of the BCGs are offset
from the true centers of the halos~\citep{George2012}.  To mitigate such a
miscentering effect, we calculate a projected offset $d_{\rm offset}$
between the BCG and the luminosity-weighted center for each group,
and remove the poorly-centered systems~($20\%$ of the total sample) with
$d_{\rm offset}{>}0.2\, r_h^{\rm AM}$, where $r_h^{\rm AM}$ is the
catalog-provided halo radius.

\emph{Large-scale galaxy overdensity of halos: $\dg$}---We are primarily interested in the dependence of HOD on the large-scale
overdensity environment~\citep{McEwen2018,Wibking2019,Contreras2021}.  For
each halo\footnote{Unless otherwise noted, we use halos and groups
interchangeably throughout the paper.} in our fiducial DESI group sample,
we measure the galaxy overdensity $\dg{=}n_g^{\rm sh}/\bar{n}_g{-}1$ within
a cylindrical shell centered on the halo, where $n_g^{\rm sh}$ is the
galaxy number density within the shell. The cylindrical shell has a radial
extent of $5$--$15\,\hmpc$~(width of $10\,\hmpc$) and a height of
$3000\,\mathrm{km/s}$ along the line-of-sight~(LOS). The exclusion of the
inner $5\,\mpch$ region is for removing the contribution of satellite
galaxies to $\dg$, but our conclusion does not depend on the inner cutoff
radius. We adopt the LSS weights when counting galaxies and take into
account the survey geometry~(including survey boundary and masks) when
computing $\dg$~\citep{Salcedo2022}, and use the rank order of $\dg$~(i.e.,
$\tdg{\in}[0,1]$), as the relevant secondary halo property for detecting
GAB.  We then divide the fiducial group sample into two equal-size
subsamples of low-$\dg$~($\tdg{<}0.5$) and
high-$\dg$~($\tdg{\geqslant}0.5$) halos for the GAB analysis. Within each
$\dg$ split, we further divide the halos into five mass bins according to
their $\mham$, with the bin edges at $\log\mham{=}[13, 13.3, 13.6, 13.9,
14.2, 15.0]$.  We emphasize that the $\mham$ values are used solely for
binning in our analysis, while the true mass $M_h$ of each $\mham$ bin will
be directly measured from weak lensing.  Consequently, our final
measurement of $\ngalmh$ is independent of the binning scheme.

\begin{figure}[t!]
     \centering
     \includegraphics[width=.9\columnwidth]{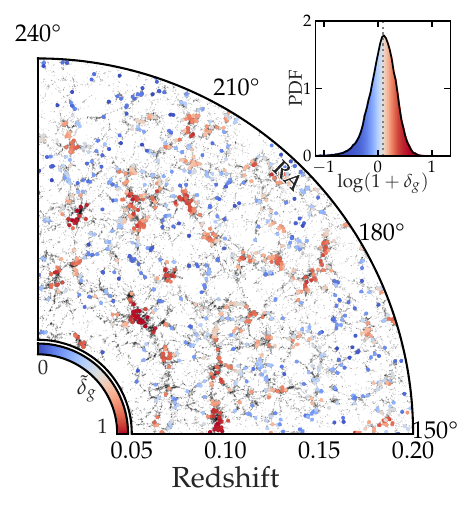}
     \caption{Distribution of DESI groups~(circles) in a redshift
     wedge~($|\rm Dec| {<} 1^\circ$) within the DESI DR1 volume.  Each
     group is color-coded by its rank-order in galaxy overdensity $\tdg$,
     according to the curved colorbar in the bottom left. Gray dots show
     the distribution of DESI BGS galaxies. The upper right panel shows the
     PDF of $\dg$, with the vertical dotted line indicating the median used
     for splitting the groups into low- and high-$\dg$ subsamples.
     \label{fig:sample}}
\end{figure}

Figure~\ref{fig:sample} compares the distributions of high-$\dg$~(redder
circles) and low-$\dg$ halos~(bluer circles) within the large-scale
structure of BGS galaxies~(gray dots), inside a redshift wedge of $|\rm
Dec| {<} 1^\circ$.  The different shades of the circles are color-coded by
$\tdg$, as indicated by the curved colorbar in the bottom left. The
probability density distribution~(PDF) of $\dg$ is shown in the upper right
panel, with the median value indicated by the dotted vertical line.
As expected, the high-$\dg$ halos exhibit a strong clustering pattern
within the densest regions of the cosmic web, whereas the low-$\dg$ halos
are preferentially found in the field and even in the voids.

\emph{Halo-galaxy cross-correlation and weak lensing}---As outlined in the Introduction, our direct method of detecting GAB
requires measuring the halo-galaxy cross-correlation function $\wphg$ and
the weak lensing profile $\dsh$.  We measure $\wphg$ by integrating the 2D
redshift-space correlation function $\xihg^{rs}$ along the LOS
\begin{equation}
    \wphg(r_p)=2\int_0^{r_\pi^{\mathrm{max}}}\xihg^{rs}(r_p, r_\pi)\,\mathrm{d}r_\pi,
    \label{eqn:wphgrs}
\end{equation}
where $r_p$ and $r_\pi$ are the projected and LOS pair separations,
respectively. We compute $\xihg^{rs}$ using the Landy-Szalay
estimator~\citep{Landy1993}, and set the integration limit $\pi_{\rm max}$
to be $200\,\mpch$ when calculating $\wphg$.

For measuring $\dsh(r_p)$, we use the public shear
catalog\footnote{\url{https://www.andrew.cmu.edu/user/rmandelb/data.html}}
based on the SDSS DR7~\citep{SDSSDR7}, which is described in detail
in Ref.~\cite{Reyes2012} and systematically tested
in Refs.~\cite{Mandelbaum2012,Mandelbaum2013}.  The shapes of galaxies are
measured using the re-Gaussianization method~\citep{Hirata2003} and the
photometric redshifts of the sources are estimated using Zurich
Extragalactic Bayesian Redshift Analyzer~(ZEBRA; \citep{Feldmann2006}).
The $\dsh$ profiles are measured from the shear catalog by following the
procedures in Ref.~\cite{Singh2020}. The covariance matrices of $\wphg$ and
$\dsh$ are both estimated by applying the jackknife resampling technique
using 200 equal-area, spatially contiguous subregions within the DESI Y1
footprint.

\emph{Direct measurement of $\ngalmhdg$}---For any $\mham$ bin in the high- or low-$\dg$ subsample, we first measure
the average halo mass $M_h$~(hence radius $r_h$) from $\dsh(r_p)$,
following the same method described in Ref.~\cite{Zu2021bcg}.  We briefly
describe the method below and refer readers to Ref.~\cite{Zu2021bcg} for
technical details. In the absence of miscentering, the weak lensing profile
can be modelled as
\begin{equation}
    \dsh (r_p | M_h)  = \frac{2}{r^2_p}\int_0^{r_p}   r_p^\prime
    \Sigma(r_p^\prime | M_h)\,\mathrm{d}r_p^\prime - \Sigma(r_p| M_h),
\end{equation}
where $\Sigma(r_p | M_h)$ is the surface mass density profile of halos with
mass $M_h$. We then compute $\Sigma(r_p | M_h)$ as
\begin{equation}
    \Sigma(r_p | M_h) =
    2\rho_m \int_{0}^{+\infty}\xi_{hm}\left(r{=}\sqrt{r_p^2+r_\pi^2}\Big|M_h\right)\,\mathrm{d}r_\pi.
\end{equation}
where $\xihm(r|M_h)$ is the isotropic halo-mass cross-correlation function.
Our analytic model of $\xihm(r)$ is presented in the End Matter.  To
further account for the miscentering effect, we adopt the flexible
miscentering model in Ref.~\cite{Zu2021bcg}, which allows an arbitrary
fraction of miscentered halos that are offset from the halo center by a
shape-2 Gamma distribution~\citep{Zhang2019}.

With $M_h$ measured from weak lensing,
the mean galaxy occupation for a set of perfectly-centered halos
with exactly the same mass $M_h$~(hence the same $r_h$) can be computed by
\begin{equation}
    \ngal^\prime(M_h)=n_{g} \int_0^{r_{h}} 4\pi r^2 \xihg(r) \,\mathrm{d}r
    + n_g V_{\mathrm{sph}}(r_h),
    \label{eqn:ngalprime}
\end{equation}
where $\xihg(r)$ is the isotropic halo-galaxy cross-correlation function
(see End Matter), $n_g$ is the mean
galaxy number density,
and $V_{\mathrm{sph}}(r_h)$ is the volume of the spherical halo. Following
Ref.~\cite{Zu2021mgii}, we infer $\xihg(r)$ via the deprojection of the
observed $\wphg(r_p)$ assuming
\begin{equation}
    \wphg(r_p)=2
    \int_0^{r_\pi^{\mathrm{max}}}\xihg\left(r{=}\sqrt{r_p^2+r_\pi^2}\right)\,\mathrm{d}r_\pi.
    \label{eqn:wphg}
\end{equation}

\begin{figure*}[ht!]
     \centering
     \includegraphics[width=\textwidth]{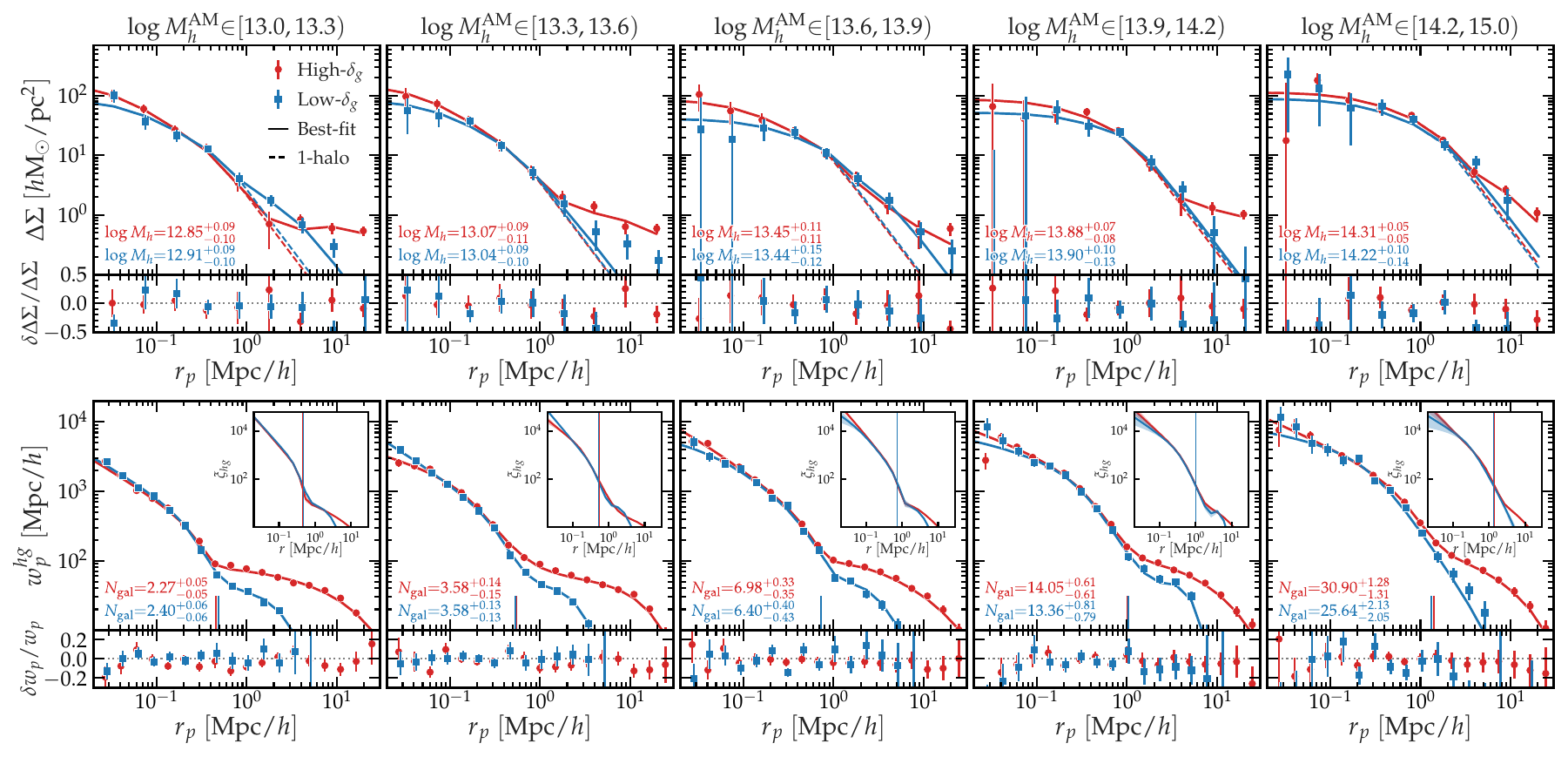}
     \caption{{\it Top}: Weak lensing profiles $\dsh$ of the
     high-$\dg$~(red) and low-$\dg$~(blue) groups in five different $\mham$
     bins. In each panel, symbols with errorbars and solid curves of
     corresponding colors are the measurements and best-fitting model
     predictions, respectively. Dashed curves indicate the 1-halo
     contribution to the model, with the corresponding halo mass
     constraints listed in the bottom left corner. Fractional differences
     between the measured and predicted profiles are shown in the bottom
     subpanel.  {\it Bottom}: Similar as the top but for the projected
     halo-galaxy cross-correlation functions $\wphg$.  In each panel,
     constraints on the occupation numbers are listed in the bottom left
     corner, while the inset panel shows the isotropic halo-galaxy
     cross-correlation function $\xihg$ predicted by the best-fitting
     model. The vertical colored lines mark the average halo radii of the
     two subsamples inferred from the weak lensing masses.
     \label{fig:obsdswp}}
\end{figure*}

Armed with the inferred $\xihg$, we compute a preliminary occupation number
$\ngal^\prime$ using Equation~\ref{eqn:ngalprime}, for the ideal case in
which all halos are correctly-centered spheres with the same radius.
However, the observed groups could be miscentered and have a ${\sim}0.2$
dex scatter between $\mham$ and $M_h$~\citep{Yang2021}. These deviations
from the ideal case will cause the satellite distribution to extend beyond
the average halo radius $r_h$. Consequently, $\ngal^\prime$ will be
underestimated compared to the true occupation number $\ngal$. To quantify
such a systematic discrepancy, we define a ``galaxy occupation bias''
parameter $q$, so that
\begin{equation}
    \ngal = (1+q) \cdot \ngal^\prime.
    \label{eqn:q}
\end{equation}
For detecting GAB, it is critical that $q$ does not depend on $\dg$,
otherwise the occupation bias will masquerade as a false GAB signal.
Through a high-fidelity DESI-like mock test, we demonstrate that $q$ is
independent of $\dg$, and the fiducial value of $q$ is equal to $0.102$.
Thus, we compute the final galaxy occupation numbers as $\ngal(M_h | \dg)
{=} 1.102 {\times} \ngal^\prime(M_h | \dg)$. The details of the mock test
are presented in the End Matter.

Figure~\ref{fig:obsdswp} presents an overview of our direct measurements of
$\ngalmhdg$ for the five different $\mham$ bins.  The top and bottom rows
show the $M_h$ measurements from $\dsh(r_p)$ profiles and $\ngal$
measurements from $\wphg(r_p)$ profiles, respectively.  In each panel, red
circles and blue squares with errorbars are the profiles measured for the
halos in high- and low-$\dg$ environments, respectively.  Solid curves with
matching colors show the predictions from the posterior median models.  The
fractional differences between the measurements and predictions are shown
in the bottom subpanels. In the top row, the $1{-}\sigma$ constraints on
the respective halo masses are listed in the bottom left corner of the main
panel. We also show the $\dsh$ contributions from within the halo radius
(a.k.a., the 1-halo term) as dashed curves.  In the bottom row, we show the
posterior median models of $\xihg(r)$ in the inset panels and list the
$1{-}\sigma$ constraints on $\ngal$ in the bottom left of the main panel.
Vertical lines in both the main and inset panels indicate the average halo
radii inferred from weak lensing.

\begin{figure}[t!]
     \centering
     \includegraphics[width=\linewidth]{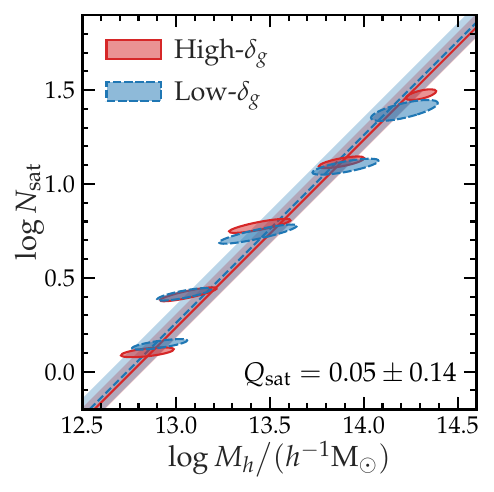}
     \caption{Satellite occupation number as a function of halo mass in
	 high-$\dg$~(red) and low-$\dg$~(blue) environment.
     Each ellipse represents the $68\%$ confidence region of the
     joint constraint on the $\nsat$ vs. $M_h$ plane for one group
     subsample. Red and blue lines with shaded bands are the
     respective predictions from our best-fitting satellite occupation
     model with galaxy assembly bias~($\qparres$).
     \label{fig:Nsat_dg}}
\end{figure}

\emph{Galaxy assembly bias constraint: $\qpar$}---We now search for evidence of GAB by comparing the $\ngalmhdg$ measurements
between the high-$\dg$~(red) and low-$\dg$~(blue) subsamples in
Figure~\ref{fig:Nsat_dg}. Since the central galaxies of the groups used in
our analysis are all brighter than $0.4 L_*$
(the luminosity threshold of our galaxy sample),
the central galaxy occupation
number is strictly unity; we thus focus on the GAB constraint of satellite
galaxies using the satellite occupation number $\nsatmhdg{=}\ngalmhdg{-}1$.
Each pair of ellipses in Figure~\ref{fig:Nsat_dg} represents the $68\%$
confidence regions of our joint constraint of $\nsat$ and $M_h$, for halos
in separate $\dg$ environments but within the same $\mham$ mass bin.  The
positive correlation between $\nsat$ and $M_h$ is caused by the fact that a
higher $M_h$ would require integrating $\xihg(r)$ to a larger $r_h$,
yielding a higher $\nsat$. Note that since each pair of ellipses
corresponds to halos of different $M_h$, we cannot measure GAB by comparing
the $\nsat$ values within the pair.

In order to detect GAB self-consistently, we employ a unified GAB model to
fit to both sets of $\nsatmhdg$ measurements simultaneously.
We assume the HOD at any given $\dg$ follows a power-law
\begin{equation}
     \nsatmhdg=\left[\frac{M_h}{M_1(\dg)}\right]^\alpha,
\end{equation}
where $M_1$ is the typical mass of a halo that can host one galaxy at any
given $\dg$ environment.  Following Ref.~\cite{Salcedo2022}, we model GAB as
the modulation of $M_1$ by the overdensity rank-order $\tdg$,
\begin{equation}
    \log M_1(\dg)=\log {M}_{1}^{*}+\qpar(\tdg-0.5),
\end{equation}
where $M_{1}^*$ is the $M_1$ at the median overdensity $\tdg=0.5$, and
$\qpar$ is the GAB parameter~($\qpar{=}0$ corresponds to zero GAB).  A
positive $\qpar$ means that halos in high-$\dg$ environments host fewer
number of satellites than ones in low-$\dg$ environments.  During the fit, we
set $\tdg$ to be $0.25$ and $0.75$ for the low- and high-$\dg$ subsamples,
respectively, and assume a fixed value of $\alpha{=}1$~\citep{Zehavi2011}.

Finally, taking into account the full covariance of all $(\nsat, M_h)$
pairs, we obtain a GAB constraint of $\qparres$.  The best-fitting model
fits and the associated $1\sigma$ uncertainties are shown in
Figure~\ref{fig:Nsat_dg} as the two lines and shaded regions,
respectively. Despite the slight preference towards a positive $\qpar$,
the two power-law fits are consistent with each other.

The projection effect may introduce a $\dg$-dependent bias in the
measurement of $M_h$, we test the impact of such an effect on our GAB
constraint by assuming the high-$\dg$ halos are subject to an additional
5\% bias in their estimated values of $M_h$, according to the mock
experiment in Ref.~\cite{Wu2022}. As a result, the GAB constraint shifts
slightly to $\qpar{=}-0.01{\pm}0.14$ and becomes even more consistent with
zero. Therefore, our direct GAB measurement suggests that there is no
statistically significant dependence of the galaxy occupation number on the
large-scale overdensity, at least for DESI galaxies brighter than $0.4 L_*$
and halos with $\log M_h > 12.9$.

\emph{Conclusions}---In this Letter, we develop a novel method to directly measure the HOD of
galaxies in DESI groups, $\ngal(M_h)$, by combining the group-galaxy
cross-correlation functions with the halo mass measurements from weak
lensing. By examining the $\ngal(M_h)$ measurements for halos within
different environments, characterized by their large-scale galaxy
overdensities $\dg$, we are able to directly measure the level of galaxy
assembly bias with the DESI DR1 data. In particular, we apply our direct
method to measure $\ngalmhdg$ using a volume-limited~($0.05{<}z{<}0.2$ and
$M_r{\leqslant}{-}19.68$) galaxy sample from the DESI BGS and the
state-of-the-art halo-based group catalog from \citet{Yang2021}. Focusing
on the satellite galaxies, we obtain a stringent constraint on the
satellite assembly bias parameter as $\qparres$, suggesting zero galaxy
assembly bias within our DESI galaxy sample~(${\geqslant}0.4 L_*$) inside
group-size halos~($\log M_h {>} 12.9$).

Our result is consistent with the comprehensive indirect constraint based
on the SDSS data~($\qpar{=}0.09{\pm}{0.10}$; \citep[][]{Salcedo2022}),
despite the slight difference in sample definition ($L/L_*{\geqslant}0.4$
vs. $M/M_*{\geqslant}0.6$). Recent studies \citep{Alam2019,Dominguez2025}
also found little dependence of HOD on the cosmic web environment using the
SDSS data. The combination of these independent constraints reveals a
remarkably simple picture of galaxy-halo connection in the local Universe.
Despite the visually striking difference between the large-scale
environments of the two halo subsamples~(as seen in
Figure~\ref{fig:sample}), the number of BGS galaxies hosted by the halos is
almost entirely determined by the halo mass~\citep{White1997}.

However, our direct constraint is in tension with the significant amount of
GAB predicted by various empirical models of galaxy formation. For
example, Ref.~\cite{Xu2021} found $\qpar{=}{-}0.25$ in a semi-analytic
model, using a mock galaxy sample that has the same number density as ours.
It is unclear whether the apparent over-prediction of GAB by empirical
models is an artifact of the subgrid physics, or strong GAB is strictly a
phenomenon in the lower-mass regime~(e.g., ${<}0.4L_*$ or $\log
M_h{<}12.9$). This issue can be addressed by applying our direct method to
the fainter galaxy samples within DESI BGS using future data releases.

Beyond DESI BGS, our method can be straightforwardly extended to other
populations of tracer galaxies in DESI, especially the LRGs~\citep{LRG2023}
at the higher redshifts, as well as the dense samples of emission line
galaxies in the upcoming Prime Focus Spectrograph~\citep{PFS2014} survey.
These galaxies have complex color selections and may therefore have
different levels of GAB compared to the luminosity-limited sample studied
in this work (e.g.,
\cite{Montero-Dorta2017,Niemiec2018,Obuljen2020,Yuan2021,Oyarzun2024}).  Our current
constraint on GAB is fundamentally limited by the uncertainties in the halo
mass and the availability of high-quality group catalogs, both of which
will be significantly improved in the era of the Vera C. Rubin
Observatory~\citep{LSST2019}, {\it Euclid} mission~\citep{Euclid2025},
Nancy Grace Roman Space Telescope~\citep{Roman2019}, and China Space
Station Telescope~\citep{CSST2025}.

\emph{Acknowledgments}---We thank Sownak Bose, Andrew Hearin, Rachel
Mandelbaum, Shun Saito, and Tae-hyeon Shin for helpful comments and
discussions. This work is supported by the National Key Basic Research and
Development Program of China (No.  2023YFA1607800, 2023YFA1607804), the
National Science Foundation of China (12173024), and the China Manned Space
Program (No. CMS-CSST-2025-A04).  Z.S. is supported by T.D. Lee
scholarship.  Y.Z. acknowledges the generous sponsorship from Yangyang
Development Fund.  Y.Z. thanks Cathy Huang for her hospitality at the
Zhangjiang High-tech Park.

This material is based upon work supported by the U.S. Department of Energy
(DOE), Office of Science, Office of High-Energy Physics, under Contract
No. DE–AC02–05CH11231, and by the National Energy Research Scientific Computing Center,
a DOE Office of Science User Facility under the same contract.
Additional support for DESI was provided by the U.S. National Science Foundation (NSF),
Division of Astronomical Sciences under Contract No. AST-0950945 to the NSF’s
National Optical-Infrared Astronomy Research Laboratory; the Science and Technology
Facilities Council of the United Kingdom; the Gordon and Betty Moore Foundation;
the Heising-Simons Foundation;
the French Alternative Energies and Atomic Energy Commission (CEA);
the National Council of Humanities, Science and Technology of Mexico (CONAHCYT);
the Ministry of Science, Innovation and Universities of Spain (MICIU/AEI/10.13039/501100011033),
and by the DESI Member Institutions: \url{https://www.desi.lbl.gov/collaborating-institutions}.
Any opinions, findings, and conclusions or recommendations expressed in this
material are those of the author(s) and do not necessarily reflect the views of
the U. S. National Science Foundation, the U. S. Department of Energy, or any of
the listed funding agencies.

The authors are honored to be permitted to conduct scientific research on
I'oligam Du'ag (Kitt Peak), a mountain with particular significance to the Tohono O’odham Nation.

This work has used the following python packages: \texttt{NumPy}, \texttt{SciPy}, \texttt{Astropy}, \texttt{dsigma} \citep{numpy2020, scipy2020, astropy2013,astropy2018,astropy2022, dsigma2022}.

\emph{Data availability}---All data points in the figures are available in the DESI organization repository of Zenodo \footnote{\url{https://doi.org/10.5281/zenodo.17387576}}.

\bibliography{gab}{}
\bibliographystyle{apsrev4-2}

\clearpage
\section*{End Matter}
\emph{Models of $\xihm(r)$ and $\xihg(r)$}---We model the isotropic cross-correlation function between halos and
secondary tracers $x$ ($x$ can be either $g$ or $m$), $\xihx$,
as the sum of two components:
\begin{equation}
     \xihx(r)=\xi^{\rm in}_{hx}(r)+\xi^{\rm out}_{hx}(r),
\end{equation}
where the two terms on the right-hand side describe the tracer distribution
within and beyond halo radius, respectively.

For the outer profile $\xihx^{\rm out}$ we develop a model
with five parameters $\bm{\theta}_{\rm out}{=}\{ B, r_0, \delta, m, \sigma_{\rm ex}\}$, modified from the one proposed by Ref.~\cite{Diemer2014},
\begin{equation}
    \xihx^{\rm out}(r | \bm{\theta}_{\rm out})=B[1+(r/r_0)^{\delta}]^{-1}
     e^{-mr/r_0}\cdot f_{\rm ex}(r).
\end{equation}
Here the exponential term is necessary for modeling the profile variation with
$\dg$, and $f_{\rm ex}$ is introduced by Ref.~\cite{Zu2021mgii} to describe the halo exclusion effect,
\begin{equation}
    f_{\rm ex}(r)= \left(1+{\rm erf}\left[{(r-r_0)}/{\sigma_{\rm
     ex}}\right]\right)/2.
\end{equation}

For the inner profiles, we model $\xihm^{\rm in}(r)$ and
$\xihg^{\rm in}(r)$ separately since the spatial distribution
of galaxies may not follow that of dark matter \cite{Budzynski2012}.
Following Ref.~\cite{Zu2021bcg}, we model $\xihm^{\rm in}(r)$
as a Navarro-Frenk-White profile~(NFW; \citep{Navarro1997}) truncated
at $r_h$.
The halo-galaxy cross-correlation function $\xi^{\rm in}_{hg}(r)$ is modelled
as a truncated generalized NFW profile~(gNFW; \citep[][]{Zhao1996}) with six
parameters $\bm{\theta}_{\rm in} {=} \{A, r_s, \alpha, \beta, \gamma, \sigma_t\}$,
\begin{equation}
     \xi^{\rm in}_{hg}(r | \bm{\theta}_{\rm in})= \rho_{\rm gNFW}(r)\cdot f_t(r),
\end{equation}
where
\begin{equation}
     \rho_{\rm gNFW}(r) =
     A\left(r/\rs\right)^{-\gamma}
     \left[1+\left(r/\rs\right)^\alpha\right]^{(\gamma-\beta)/\alpha},
\end{equation}
and
\begin{equation}
     f_{t}(r) = \left[1-{\rm erf}\left((r-r_{h})/{\sigma_{t}}\right)
     \right]/2.
\end{equation}

\emph{Mock Calibration of ``Occupation Bias'' $q$}---In order to calibrate the value of $q$ and more importantly, ascertain
whether $q$ varies with $\dg$, we perform a comprehensive mock experiment
using the simulated halo catalog from the $z{=}0$ snapshot of the
\texttt{ELUCID} simulation (same as that used by Ref.~\citep{Salcedo2022}). A
mock galaxy catalog is produced by populating the \texttt{ELUCID} halos
with the best-fitting \texttt{iHOD} parameters derived in Ref.~\cite{Zu2015},
assuming an NFW distribution for the satellite galaxies. We select from the
\texttt{iHOD} mock catalog a stellar-mass limited mock sample that has the
same $n_g$ as our DESI sample.  For the mock halo catalog, we generate mass
observables $M_h^{\rm mock}$ by adding a Gaussian random noise of
$0.2$ dex to the true halo mass.  To mimic miscentering, we shift the centers of $30\%$ of the halos by $d_{\rm offset}{=}x r_h$, where $x$ follows the shape-2 Gamma distribution
with a scale parameter $0.2$.  We select halos with $13 {<} M_h^{\rm mock}
{<} 13.3$ into our mock halo sample. The large-scale galaxy overdensity
$\dg$ is then calculated for each halo using the mock galaxy sample.

\begin{figure*} \centering
    \includegraphics[width=\textwidth]{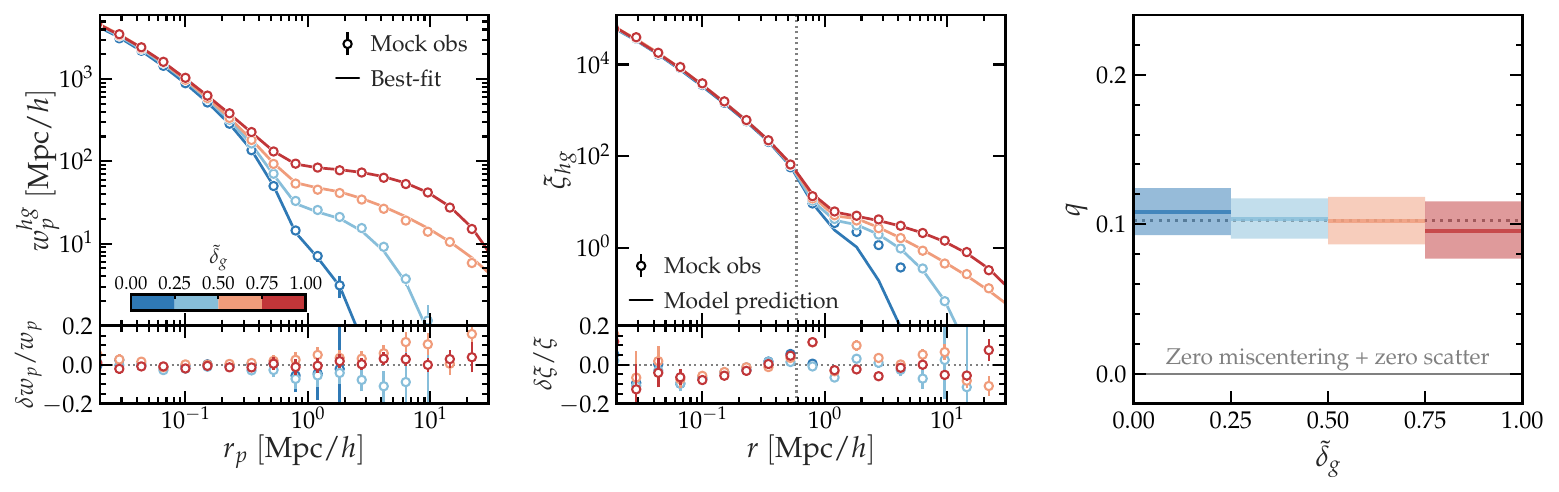}
    \caption{Mock experiment testing the sensitivity of ``galaxy occupation
    bias'' $q$~(fractional difference between the inferred and true
    occupation number; see Equation~\ref{eqn:q}) to large-scale overdensity
    $\dg$.  {\it Left}: Projected halo-galaxy cross-correlations $\wphg$
    measured from the mock data~(circles with errorbars) and predicted by
    the best-fitting model~(curves) for halos in four quartiles of $\dg$,
    color-coded by the colorbar in the bottom left. Bottom subpanel shows
    the fractional differences between the mock measurements and
    predictions.  {\it Middle}: Comparison between the isotropic
    cross-correlation functions $\xihg$ measured from the mock~(circles)
    and inferred from the fits to $\wphg$~(curves).  Vertical dotted line
    marks the average halo radius.  {\it Right}: Occupation bias $q$ as a
    function of overdensity rank-order $\tdg$. Thick horizontal lines with
    shaded bands indicate the $1\sigma$ measurements of $q$ in the four
    quartiles of $\dg$.  Dotted horizontal line denotes the mean value of
    $q$, while solid horizontal line marks the $q{=}0$ case when there is
    no miscentering or scatter in the mass-observable relation.
\label{fig:mock}}
\end{figure*}

Using the mock halo and galaxy samples, we compute the $\wphg(r_p)$
profiles for halos in four quartile of $\dg$, and then infer the
$\xihg(r_p)$ profiles using the method described before.
The results are shown in Figure~\ref{fig:mock}.  In the left panel, the
$\wphg$ profiles measured for mock halos in four $\dg$-quartiles (circles
with errorbars) are well described by predictions from the posterior mean
models~(solid curves). The circles and curves are color-coded by $\dg$,
according to the colorbar in the bottom left. The middle panel compares the
corresponding $\xihg$ predictions~(curves) with the direct measurements
from the simulation~(circles with errorbars). The model curves recover the
simulation results to within 5\% below the halo radius~(bottom subpanel),
which is indicated by the vertical dotted line at around $0.6\,\hmpc$.  The
$\xihg$ discrepancies on scales beyond $r_h$~(e.g., in the lowest $\dg$
quartile) do not affect the $\ngal^\prime$ measurements. At last, we use
Equation~\ref{eqn:ngalprime} to calculate $\ngal^\prime(\dg)$ for the mock
halos in different $\dg$ quartiles, by integrating the model $\xihg$ curves
from the middle panel.

The right panel of Figure~\ref{fig:mock} presents the dependence of $q$ on
the $\dg$ quartile, with the widths of the horizontal bars indicating the
measurement uncertainties of $q$. The gray horizontal line denotes
$q{=}0$ for the ideal case with zero miscentering and zero scatter, while the
dotted horizontal line shows the average of the four quartiles
$\bar{q}{=}0.102$. Therefore, for a reasonable level of miscentering and
a $0.2$-dex scatter in the $\mham{-}M_h$ relation, we expect an
occupation bias around ${\sim}10\%$ level. More important, the individual
$q$ measurements for the four $\dg$ quartiles are all consistent with
$\bar{q}$ within the errorbars --- we do not detect any dependence of $q$
on the overdensity environment.


\end{document}